\documentclass[twocolumn,showpacs,preprintnumbers,amsmath,amssymb,superscriptaddress,10pt,aps,prb]{revtex4-1}
\usepackage{epsfig,amsopn}
\usepackage{graphicx}
\usepackage{epstopdf}
\usepackage{amsmath,amssymb}
\usepackage{amsthm}
\usepackage{enumerate}
\usepackage{xcolor}

\begin{document}
	\title[Quantum transport simulation of non-local response in Weyl semimetals]{Quantum transport simulation of non-local response in Weyl semimetals}
	\author{Keshav Pareek and Arijit Kundu}
	\affiliation{Department of Physics, Indian Institute of Technology Kanpur, Kanpur 208016, India}

\begin{abstract}
We numerically study the non-local transport signature in Weyl semimetal as a test for interconnectedness of the surface states, using a recursive Green's function method. We drive a current using two leads connected on the same surface (top surface) and apply a magnetic field throughout the system, perpendicular to the surface. We find that this results in a current flowing on the other surface in the direction opposite to the direction of the current on the top surface and we comment on the viability of observing such an effect in experiment. The recursive Green's function method we employ is exact and provides us with the Green's functions of the two surfaces as well as their connecting elements, which can be applied also for other numerical simulations where the effect of surface to surface transport is important.
\end{abstract}

\maketitle

\section{Introduction}
Weyl semimetals (WSM)~\cite{Vishwanath2011,Burkov2011a,Burkov2011b,Zyuzin2012a,Hosur2012,rmp} are three dimensional topological systems with gap-less points in the spectrum near Fermi energy, called Weyl nodes. Quasi-particle excitations near these momentum points follow a relativistic Weyl equation and behaves like Weyl fermions. Such \textit{Weyl nodes} always appear in pairs of opposite \textit{chirality}~\cite{Nielsen1981}. The low-energy excitations of the system are expected to show a number of novel behaviors: such as extremely high mobility and negative longitudinal magneto-resistance. Possibility of observing novel physical phenomena as well as the promise WSMs present for futuristic applications created an enormous excitement in the condensed matter and material physics community for the last few years, yielding a number of achievements in terms of theoretical studies~\cite{Vazifeh2013,Turner2013,Son2013,Biswas2013,Hosur2013,Burkov2014,Gorbar2014,Uchida2014, Khanna2014,Ominato2014,Sbierski2014,Burkov2015a,Burkov2015b,Goswami2015,nonlocal1, Khanna2016,Behrends2016,Rao2016,Baireuther2016a,Tao2016,Marra2016,Li2016,Baireuther2016b,Madsen2016,Obrien2017,Khanna2017,Bovenzi2017,strainWSM} and experimental verification~\cite{WSMARPES,Xu2015a,Xu2015b,Lv2015a,Lv2015b,Lu2015,Jia2016,aag}.

\begin{figure}[t]
	\centering
	\includegraphics[width=0.4\textwidth]{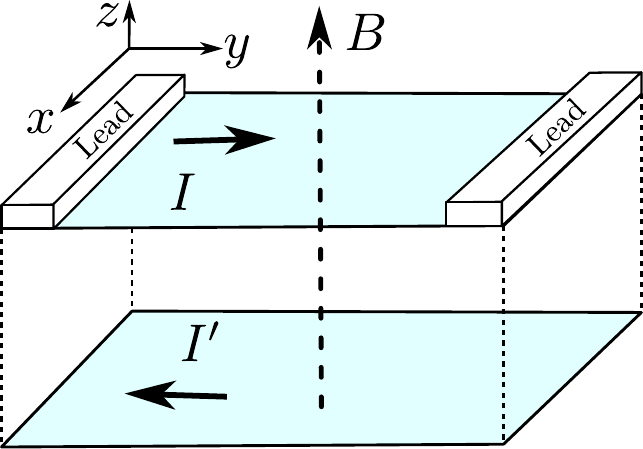}\\
	\caption{The setup for the quantum transport simulation. The leads, distance $L_y$ apart, are attached at the surface of a WSM of finite height $L_z$. In our results, we keep the translational invariance of the transverse direction, $x$, intact and a perpendicular magnetic field of strength $B$ is applied.}\label{fig:setup}
\end{figure}

A novel aspect of the physics of WSM is the non-local electro-magnetic response in bulk and surface transport~\cite{nonlocal1,nonlocal2,nonlocal3}. The surface states of Weyl semimetals, often called \textit{Fermi arc}, connect the two Weyl nodes of opposite chiralities~\cite{FA}. In presence of a magnetic field, Landau levels are formed with dispersion along the direction parallel to field. The low energy bulk Landau levels and Fermi arcs intermix to form closed orbits, which have been theoretically studied~\cite{paralllel} and experimentally verified~\cite{expt}. Potter et al.~\cite{potter} showed that these orbits are responsible for a ‘conveyor belt’ motion of electrons in presence of magnetic field and applied bias, imparting a non local nature to the currents at top and bottom surface. A similar ‘wormhole tunneling’ between Fermi arcs in momentum space was described by Wang et al.~\cite{warmhole} to justify the plateaus in quantum hall conductivity at the top surface, where the Weyl nodes act as wormholes in the momentum space, transporting electrons from one surface to another. In this work we aim to verify this phenomena using quantum transport calculations of surface currents, with a focus on modifying the recursive Green's function formalism capable of taking into account the connection between surface states through the bulk. Our setup, as shown in Fig.~\ref{fig:setup}, consists of a Weyl semimetal block and two leads, connected to the surface of the Weyl semimetal, which can drive a surface current $I$ on the top surface. Because the system is metallic current will flow throughout the bulk as well as the bottom surface. Now, when a magnetic field is applied in a direction as shown in the figure, we expect a reversal in the direction of the current on the bottom surface~\cite{potter}, which can serve as a signature of the non-locality of the Fermi arcs, allowing the electrons to tunnel from one surface to another through the bulk of the system. Our simulation confirms this prediction.

The paper is organized as following: in the section ~\ref{sec:2}, we introduce the model for our set up and we discuss basic bulk and surface properties of the system, connecting to the origin of the expected non-local transport signature. Next, in section~\ref{sec:bond-curr} and \ref{sec:3}, we set up the numerical method, including the method of recursive Green's function. We discuss our numerical results in section~\ref{sec:4} followed by a discussion.

\section{Model of WSM}\label{sec:2}
\begin{figure}[h]
	\centering
	\includegraphics[width=0.47\textwidth]{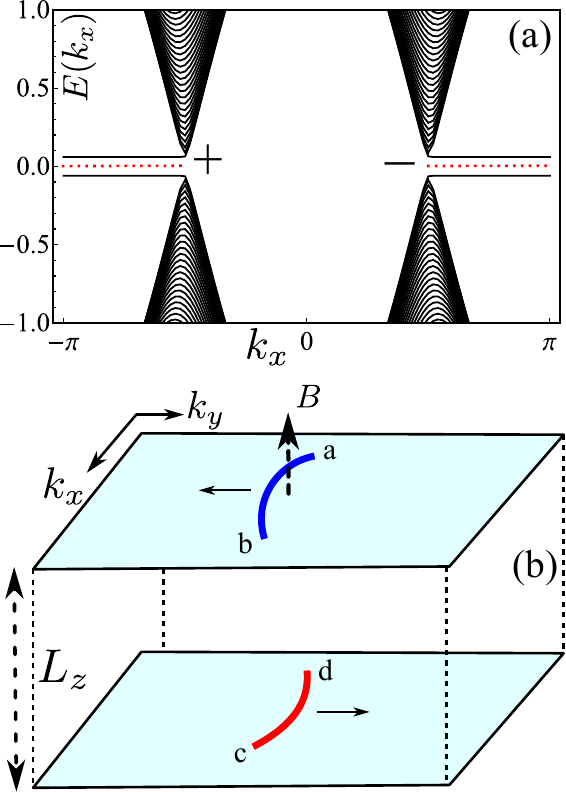}\\
	\caption{Top: the low-energy band structure of the model Eq.~(\ref{eq:H}) for a slab of finite height in $z$ direction but with translation invariance in $x$ and $y$ directions. The band-structure, as a function of $k_x$ has been plotted for $k_y = 0.05$, showing the two gap-less points (marked by $\pm$) in the bulk and the Fermi arc that joins them on the surfaces. The Fermi arcs of energies above and lower zero resides on opposite surface of the slab, whereas for $k_y=0$ and sufficiently large height of the slab, the Fermi arcs are degenerate (marked by dotted line). The parameters used are $t=t_x=2m$ and $k_0=\pi/2$. Bottom: for a finite $L_z$, schematic of Fermi arcs, joining the Weyl nodes of opposite chirality are shown in the Brillouin zone for both the surfaces. The arrow denotes for higher energy how the Fermi arc disperses as a function of momentum, giving rise to velocities, $v_i=dE/dk_i$ non-zero only in $\mp~y$ direction for upper and lower surface respectively.}\label{fig:model}
\end{figure}

We consider the following Hamiltonian of a minimal model~\cite{minmodel} for a Weyl semimetal in a cubic lattice:
\begin{eqnarray}\label{eq:H}
H(\mathbf{k}) =& \gamma (\cos k_x - \cos k_0)\sigma_0 - \left(m(2-\cos k_y - \cos k_z)\right.\nonumber \\
 & \left.+ 2t_x (\cos k_x - \cos k_0 ) \right)\sigma_1 - 2t  \sigma_2 \sin k_y \nonumber \\
 & - 2t  \sigma_3  \sin k_z.
\end{eqnarray}
Here $\sigma_{\alpha} $'s are Pauli matrices acting in the orbital space. For $\gamma<2t_x$  the bulk Hamiltonian is gap-less at two momentum points $(\pm k_0,0,0)$. Near these two momentum points the low-energy Hamiltonian follows Weyl equation with opposite chirality. For simplicity, we keep $\gamma=0$, $t_x=t$ ($t$ serves as our unit of energy) and the lattice spacing is taken to be unity. The Hamiltonian is a minimal model of \textit{time-reversal broken} Weyl semimetal where the time reversal operator is $ T= K $, with $ K $ being complex conjugation. The low-energy band-structure is shown in Fig.~\ref{fig:model}.

For a slab of finite thickness in $z$ direction, say of height $L_z$, the surface states of the Hamiltonian Eq.~(\ref{eq:H}), i.e, the Fermi arc, connects the Weyl nodes of opposite chiralities, shown in Fig.~\ref{fig:model}. These surface states have predominant velocity component along the direction perpendicular to line joining weyl nodes and the Fermi arcs on opposite surfaces have velocities in opposite directions.

The interconnectedness of the Fermi arcs is due to their topological origin and can be checked in the following way, which follows Ref.~\onlinecite{nonlocal1}. 
Consider the system as shown in Fig.~\ref{fig:setup}. The leads applies a finite bias along $y$ direction. This sets up a current along $y$ direction, with the charge carriers being predominantly from the Fermi arcs. In presence of a magnetic field along the $z$ direction, from semi-classical arguments, we can argue that these electrons will experience a Lorentz force, responsible for a change in lattice momentum of electrons-
\begin{equation}\label{eq:kdot}
\dot{\mathbf{k}} \propto \mathbf{v}\times (B\hat{z}),
\end{equation}
which is a vector tangential to the Fermi arc. Thus the electrons will slide along the Fermi arc, in the momentum space, from one chiral node to another (say, point $a$ to $b$ in Fig.~\ref{fig:model}(b)). But, due to  the finite length of the arc, once the electrons travel to $b$, they must continue to the other surface (point $c$). Now, once they start sliding to point $d$, the change of momentum, following Eq.~(\ref{eq:kdot}) will acquire the electrons with a velocity opposite to the direction of the current. This process will result in directions of the currents flowing in opposite surfaces to be opposite. We shall test this prediction in our numerical results.

\section{Formalism for bond-current}\label{sec:bond-curr}
In this section we present briefly the Green's function formalism we employ for the transport simulation. The geometry of our simulation is shown in Fig.~\ref{fig:setup}. The translation symmetry is kept intact in $x$ direction by making a suitable Gauge choice for the vector potential, $\mathbf{A} = (-By,0,0)$. We use a modified form of Landauer-Buttiker formalism to calculate the current at the top and the bottom surface between site $i$ and $i$+1 for every site, where we also sum over the contributions of modes along the transverse direction, marked by quantum number $k_x$ (we also consider $\hbar=1$ and the charge of electron $-e =-1$ throughout the text). 

For a generic non-interacting Hamiltonian 
\begin{equation}
H = \sum_{ij} \sum_{\eta,\eta'} t_{i\eta,j\eta'} c_{i\eta}^{\dagger}c_{j\eta'} + {\rm h.c.},
\end{equation}
where $i,j$ are site indices and $\eta,\eta'$ are other orbital indices. For a hopping term in the Hamiltonian, 
\begin{eqnarray}
H^{j\rightarrow i} &&= \sum_{\eta,\eta'} t_{i\eta,j\eta'} c_{i\eta}^{\dagger}c_{j\eta'} + {\rm h.c.},
\end{eqnarray}
the bond current operator is 
\begin{eqnarray}\label{eq:bondcurr}
j_{j\rightarrow i } = (-e)\left.\frac{\partial H^{j\rightarrow i}(\phi)}{\partial \phi}\right|_{\phi = 0},
\end{eqnarray}
where
\begin{eqnarray}
H^{j\rightarrow i}(\phi) = t_{i\eta,j\eta'}e^{-i\phi} c_{i\eta}^{\dagger}c_{j\eta'} + {\rm h.c.}.
\end{eqnarray}
On performing the lead average~\cite{transport}, we get the lead averaged bond current-
\begin{eqnarray}
\langle j_{j\rightarrow i } \rangle = (-e)2{\rm Im}\left[ \sum_{\eta,\eta'}t_{i\eta,j\eta'} \langle c_{i\eta}^{\dagger}(\tau)c_{j\eta'}(\tau)\rangle \right],
\end{eqnarray}
which would be independent of the time $\tau$ for a system in steady state. Going to the Fourier space and using Fermi distributions ($f^{\lambda}(\omega)$ for $\lambda^{\rm th}$ lead) of the leads~\cite{transport}, we have
\begin{align}\label{eq:bondcurrl}
\langle j_{j\rightarrow i } \rangle = -2e{\rm Im}\sum_{\eta\eta'} t_{i\eta,j\eta'} \int d\omega \sum_{\lambda} f^{\lambda}(\omega)\chi_{j\eta',i\eta}(\omega),
\end{align}
which is the quantity of interest. The correlation functions are defined as
\begin{align}
\chi_{j\eta',i\eta}(\omega) &=  \langle c_{j\eta'}^{\dagger}(\omega) c_{i\eta}(\omega)\rangle \nonumber\\
& =  (2\pi)^2\sum_{\lambda} f^{\lambda}(\omega)\left(G(\omega)V^{\lambda}G^{\dagger}(\omega)\right)_{i\eta,j\eta'},\label{eq:chi}
\end{align}
where $\lambda$ denotes the left and right leads. $G$ is the full Green's function of the system, which also includes the self-energy, $\Sigma(\omega)$, from the leads,
\begin{equation}\label{eq:GF}
G(\omega) = \left(\omega - H - i\Sigma (\omega) \right)^{-1}~,
\end{equation}
\begin{equation}\label{eq:gamma}
\Sigma_{m\eta,n\eta'}(\omega)=-i\sum_{\lambda} K_{\alpha, m\eta}^{\lambda *}[g^{\lambda}(\omega)]_{\alpha\beta}K_{\beta, n\eta'}^{\lambda}~.
\end{equation}
$ K_{\alpha, m\eta}^{\lambda}$ is the hopping amplitude from $\lambda^{th}$ lead's site $\alpha$ to the system-site $m$ and orbital $\eta$. For simplicity, we assume that the leads have a single orbital and the hopping is equally likely to either orbital of the system. Further, the density of state of leads is taken to be constant throughout the spectra (\textit{flat-band} limit). Thus, the Green's functions of leads can be represented by $g^{\lambda} = -i\pi \rho^{\lambda}$, which is independent of $\omega$ and $\rho^{\lambda}$ is the effective density of states at lead $\lambda$. Finally, we define $V^{\lambda}=K^{\lambda \dagger}\rho^{\lambda}K^{\lambda}$. The leads' chemical potentials and temperatures are encoded in the Fermi functions $f^{\lambda}(\omega)$, and a bias can be applied by using a chemical potential difference between the leads.

\begin{figure}
	\centering
	\includegraphics[width=0.45\textwidth]{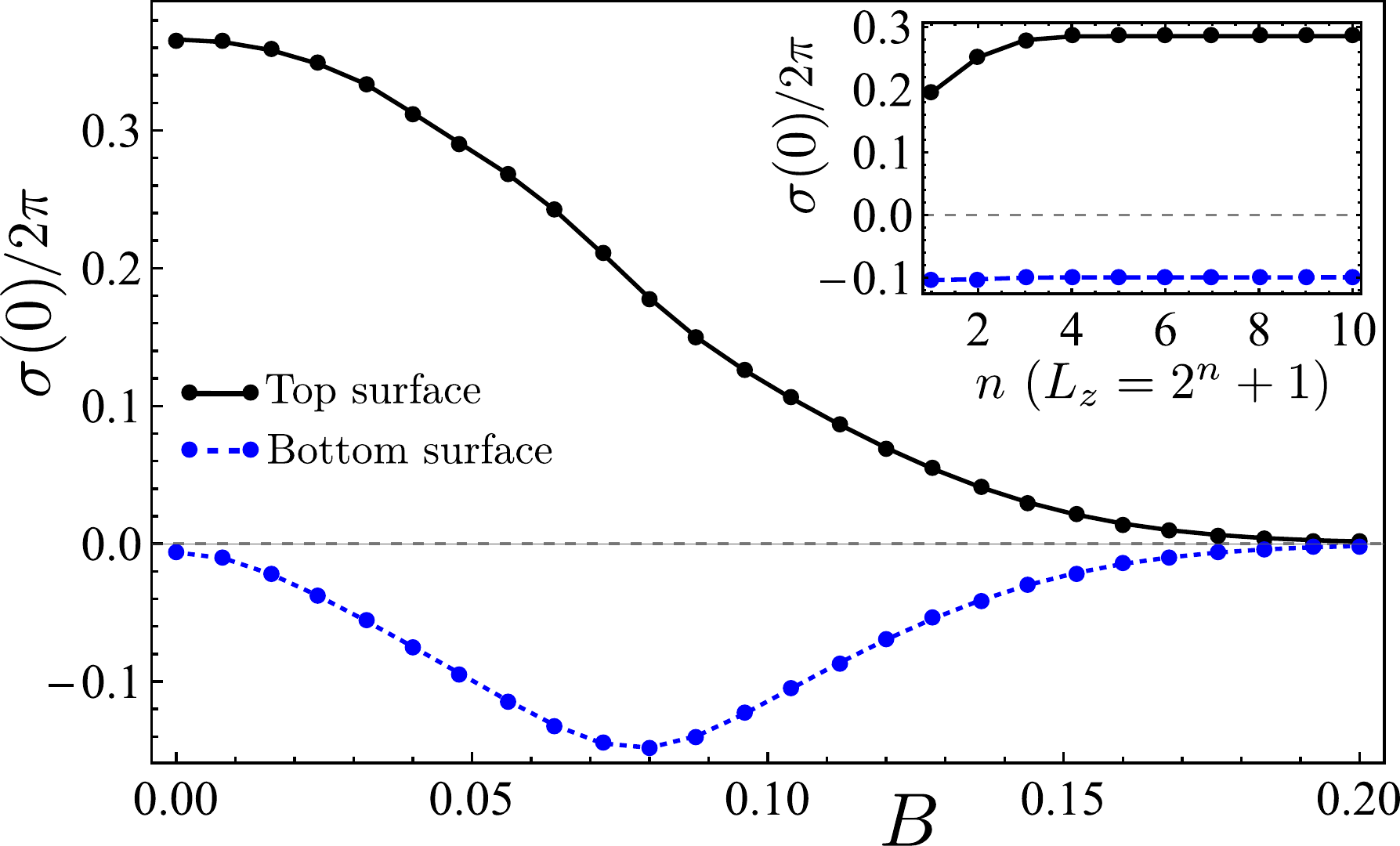}
	\caption{For small applied bias $\delta V$, the bond-current is  $\delta I = \sigma(0)\delta V$. The $\sigma(0)$ (in the unit of $e^2/h$) for the bond current along the $y$ direction, at zero temperature, for top and bottom surfaces as a function of the applied magnetic field's strength has been plotted. In the inset we plot how the currents change as a function of the height of the system, $L_z$ for a magnetic field of $B=0.05$. $L_y=40$ has been used in both figures and for the main figure we use $L_z=33$ and $k_0=\pi/2$. The leads are connected to 4 left most and 4 right most sites on the top surface, with a density of states $\rho=0.5$. The hopping amplitude from the leads' site to the system sites are unity (i.e, $=t$).}\label{fig:JvB}
\end{figure}

\begin{figure*}[ht]
	\centering
	\includegraphics[width=0.95\textwidth]{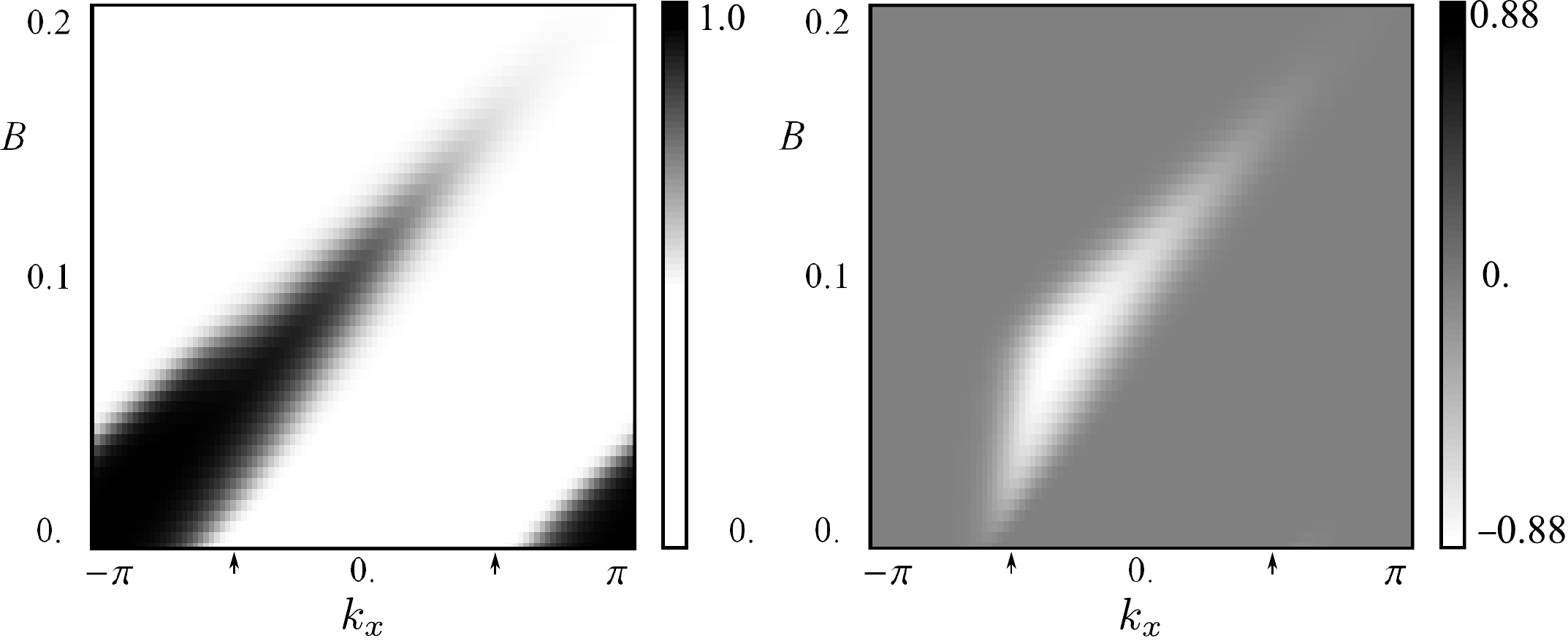}
	\caption{Differential surface conductance ($\sigma (0)$, as defined in the text) for small bias difference along the $y$ direction from site $N_y/2$ to site $N_y/2 +1$ in top (left figure) and bottom  (right figure) surfaces as a function of the applied magnetic field and the transverse momentum $k_x$. The arrows mark the positions of the Weyl nodes in the bulk in the absence of the magnetic field. Parameters used are the same as in Fig.~\ref{fig:JvB}.}\label{fig:dplot}
\end{figure*}

\section{Recursive Green's function Method}\label{sec:3}

The aforementioned Green's function for our setup can be written in a matrix form as,
\begin{eqnarray}\label{eq:gdef}
\left(\omega\mathbb{I}-\tilde{H}\right)G(\omega)=\mathbb{I},
\end{eqnarray}  
with $ \tilde{H}= H + i\Sigma_l(\omega) + i\Sigma_r(\omega)$ where, $\Sigma_l$ and $\Sigma_r$ are the self energies due to the left and the right leads. We write $\tilde{H}$ for given $k_x$ as
\begin{eqnarray}
\tilde{H} &= \left(\begin{array}{ccccc}
h_t & A & 0 & .  & 0\\
B & h & A &  . & 0\\
0 & B & h &  . & 0\\
. & . & . & .  & .\\
0 & 0 & 0 &  . & h_b\\
\end{array}\right),
\end{eqnarray}
where each of the $h$ are the Hamiltonian for a layer (given by constant $z$) and $A$ (and $B=A^{\dagger}$) is the connecting matrices from one layer to another. The self energy contribution due to leads attached at the top surface is included in $h_{t}$ and since no lead is attached to the bottom surface, $h_{b}=h$. Including the vector potential for the magnetic field by Peierls substitution, each layer's Hamiltonian, $h$ is identical, except for the top layer ($0^{\rm th}$ layer), as the self-energies in our setup only appear to the sites which are on the top surface. We consider the number of layers in $z$ direction to be $L_z =N+1$.

There are many available ways to obtain surface ($G_{00}$) and bulk Green's function recursively~\cite{rec1,rec2,rec3,rec4,rec5}, especially, when $N\rightarrow \infty$. As our requirement is to find the bond current not only on the top surface, but as well as on the bottom surface, we require not only $G_{00}$, but also $G_{0N}$ and $G_{N0}$. So, we proceed to modify the recursive Green's function method in the following way.

On multiplying the above matrices, as in Eq.~(\ref{eq:gdef}), to have
\begin{eqnarray}
(\omega-h)G_{n,\alpha} =AG_{n+1,\alpha}+BG_{n-1,\alpha}, \\      \label{eq:Gna}
(\omega-h_t)G_{0,\alpha}=\mathbb{I}\delta_{\alpha,0}+AG_{1,\alpha}, \\		\label{eq:G0a}
(\omega-h_b)G_{N,\alpha}=\mathbb{I}\delta_{\alpha,N}+BG_{N-1,\alpha},		\label{eq:GNa}
\end{eqnarray}
where $n=1,2...N-1$ and $\alpha = 0,N$. Please note that the identity matrix appearing above is for each layer and not the same as appeared in Eq.~(\ref{eq:gdef}).  Now, on substituting the values of $G_{n+1,\alpha}$ and $G_{n-1,\alpha}$ in terms of $G_{n+2,\alpha}$, $G_{n-2,\alpha}$ and $G_{n,\alpha}$ in the above equations we get, 
\begin{eqnarray}
(\omega-h^{(1)})G_{n,\alpha} =A^{(1)}G_{n+2,\alpha}+B^{(1)}G_{n-2,\alpha}, \\ \label{eq:Gna1}
(\omega-h_t^{(1)})G_{0,\alpha}=\mathbb{I}\delta_{\alpha,0}+A^{(1)}G_{2,\alpha}, \\ \label{eq:G0a1}
(\omega-h_b^{(1)})G_{N,\alpha}=\mathbb{I}\delta_{\alpha,N}+B^{(1)}G_{N-2,\alpha},  \label{eq:GNa1}
\end{eqnarray}
where $A^{(1)}$, $B^{(1)}$, $h^{(1)}$, $H_{u}^{(1)}$, $H_{d}^{(1)}$ are identified as,
\begin{eqnarray}
A^{(1)} =A^{(0)}(\omega-h^{(0)})^{-1}A^{(0)}, \\
B^{(1)} =B^{(0)}(\omega-h^{(0)})^{-1}B^{(0)}, \\
h^{(1)} =h^{(0)}+B^{(0)}(\omega-h^{(0)})A^{(0)}, \\
h_{t}^{(1)} = h_{t}^{(0)}+A^{(0)}(\omega-h^{(0)})^{-1}B^{(0)},\\
h_{b}^{(1)} = h_{b}^{(0)}+ B^{(0)}(\omega-h^{(0)})^{-1}A^{(0)}.
\end{eqnarray}
where $X^{(i)}$ denotes modified matrices after $i^{\rm th}$ iteration where $0^{\rm th}$ iteration means unmodified matrices. Now, after $k$ iterations, where $L_z= N+1 =2^k+1$ (number of layers along $z$ direction), we get,
\begin{eqnarray}
(\omega-h_{t}^{(k)})G_{0,0} =\mathbb{I}+A^{(k)}G_{N,0}, \\
(\omega-h_{b}^{(k)})G_{N,0} =B^{(k)}G_{0,0},\\
(\omega-h_{b}^{(k)})G_{N,N} =\mathbb{I}+B^{(k)}G_{0,N}, \\
(\omega-h_{t}^{(k)})G_{0,N} =A^{(k)}G_{N,N}.
\end{eqnarray}
These equations can be written in simplified form 
\begin{eqnarray}
\left( \omega\mathbb{I} -h' \right)G' = \mathbb{I}, 
\end{eqnarray}
with
\begin{eqnarray}
h' = \left(\begin{array}{cc}
h_t^{(k)} & A^{(k)} \\
B^{(k)} & h_b^{(k)} \\
\end{array}\right), \quad G' = \left(\begin{array}{cc}
G_{00} & G_{0N} \\
G_{N0} & G_{NN} \\
\end{array}\right).
\end{eqnarray}
On solving the above equation, we can find the matrices $G_{0,0}$, $G_{N,N}$, $G_{N,0}$ and $G_{0,N}$, which we use in Eq.~(\ref{eq:bondcurrl}) to obtain the bond currents in top and the bottom surfaces.

The novelty of the above method is that, effectively, it reduces the computation time for a 2D system (for each $k_x$) to an effective 1D system, evaluating both the surfaces' Green's functions as well as their connections, where \textit{no approximations has been made}. Moreover the distance between the surfaces can be controlled highly efficiently, where the separation of the surfaces grow exponentially with increasing iteration.

\section{Numerical results}\label{sec:4}
Before we present our result with a finite magnetic field, we briefly discuss the $B=0$ case. Due to the time-reversal broken nature of the Hamiltonian, even without an applied bias, the individual bond currents are non-zero (but small) on the surfaces of the WSM for momentum $k_x$ between the Weyl nodes (values of $k_x$ when the Fermi arc exists), although the net current summing over $k_x$ and the layers of $z$ vanishes. This is inherent to the model we use and can be attributed to the anomalous Hall effect~\cite{Vazifeh2013,aHall1}, giving rise to a circulating current along the edges. With an applied bias, such circulating current is lost and the direction of the net current is set by the bias. Further, as electrons from the Fermi arc are chiral on the surfaces (that is, they can only move with negative/positive velocity in top and bottom surface respectively), it is required to apply a bias accordingly to draw a surface current for small bias. This sets the sign of chemical potential of left and right leads. Since we are interested in differential conductance at $\omega=0$, a small chemical potential $\pm e\mu/2$ for left and right leads would suffice.

Now, once we apply the magnetic field, because of the mechanism of Eq.~\ref{eq:kdot}, the electrons from the Fermi arc of the top surface slide along the Fermi arc to reach the other surface. This results in a circular motion of the electrons involving the Fermi arcs of both surfaces and the bulk states. Consequently this results in opposite sign of the current flowing on the opposite surfaces. Writing the current as $J = \sigma(0)\delta V$, for a small bias $\delta V$, we show the plot of $\sigma(0)$ (in the unit of $e^2/h$) versus $B$ for both surfaces in Fig.~\ref{fig:JvB} and in Fig.~\ref{fig:dplot}, where in the later figure we plot the transverse momentum resolution of the current. Fig.~\ref{fig:JvB} clearly shows a reversal of the sign of the current in the bottom surface due to the applied magnetic field, confirming the non-local transport from one surface to another. A further interesting aspect to note that the current on the bottom surface seems to be proportional to the derivative of the top current, which is reminiscent of the relationship between the transverse and longitudinal conductance of quantum Hall systems~\cite{Halperin}, although the origin of such relation is not clear to us for our system. The positions of the initial Weyl nodes in absence of applied magnetic field are marked in Fig.~\ref{fig:dplot}, where we see that the Fermi arc states carry most of the currents on the top surface, whereas on the bottom surface the bond current is vanishingly small~\cite{note1}. With increasing magnetic field, the \textit{effective} positions of the Weyl nodes change in the momentum space and the net current on the top surface decreases, although the contributions continue to come from the Fermi arcs. Conversely, for the bottom surface, with increasing magnetic field, current continues to increase, in opposite direction to the bias. This is our main result.

We also show in the inset of Fig.~\ref{fig:JvB}, how the differential conductance on the two surfaces change with increasing the height of the sample, going as $2^n+1$ layers. Interestingly, the $\sigma(0)$ becomes independent of the height of the sample for large sample width, which is precisely the non-local character of the system. This is unrealistic from an experimental perspective and appears in our case because in our setup any momentum scattering mechanism between states of different $k_x$ is missing. In realistic situation, because of possible scattering mechanisms, the maximum width when the effect can be observed might be limited.

\begin{figure}
	\centering
	\includegraphics[width=0.45\textwidth]{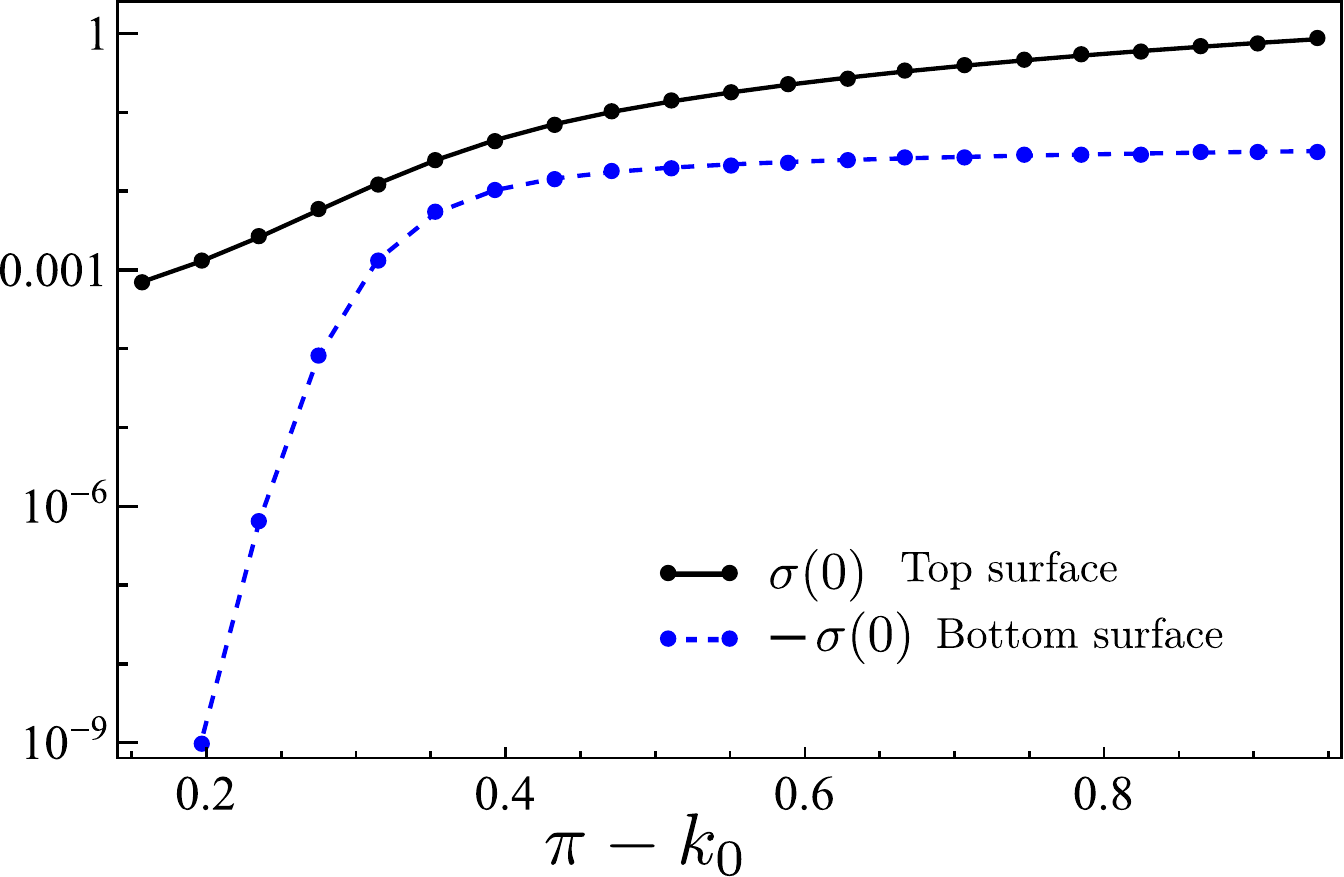}
	\caption{The dependence of the bond differential conductance on the length of the Fermi arc. Please note the log scale of the vertical axis. A larger system of size $L_x\times L_z =140\times 1025$ sites has been simulated.}\label{fig:withk0}
\end{figure}

\section{Dicussion}
Our results establishes, numerically, that the non-local transport signatures are consistent with the predictions made. Here two issues deserve special attention. First, the model we use breaks time-reversal symmetry, which results in a circulating current through the edge of the system with attached leads of the same bias and without an applied magnetic field. The crucial difference between this circulating current and the non-local transport under a finite bias and an applied magnetic field is that in the later case a net amount of current flows through the bulk of the system, resulting from the mechanism of Eq.~\ref{eq:kdot}.

In a time-reversal symmetric Hamiltonian, such circulating current without magnetic field will be absent, but there will be multiple Fermi arcs present on either surfaces. The minimal model for a time-reversal invariant WSM contains two Fermi arcs, where the electrons can move in opposite directions. Similarly, a number of approximation we have made can be lifted for more realistic setup. Namely, we used a constant density of the states of the leads; we did not allow mixing of transverse momentum in presence of the leads and we did not take into account the presence impurities. Whereas the flat-band limit of the leads' density of states is justified for small applied bias, the other constraints can be removed in our formalism, although we believe, for appropriate range of parameters, the main conclusion of our work to remain unaffected. Consideration of models of multiple Fermi arcs and more realistic leads are left for future studies.

In realistic materials, the expected length of the Fermi arcs are of the order of $0.1a^{-1}$~\cite{Huang2015b, Yang2015b}, $a$ being the lattice spacing. In our simulation, we used a larger separation of the Fermi arcs in Fig.~\ref{fig:JvB} and in Fig.~\ref{fig:dplot}. For smaller Fermi arc and smaller applied magnetic field, it is required to simulate a much larger system which is numerically difficult. The dependence on the length of the Fermi arc for the current in opposite surfaces has been plotted in Fig.~\ref{fig:withk0}, for magnetic field strength of $B=0.005\hbar^2/ea^2$ (which is of the order of tens of Tesla, considering the lattice spacing is of the order of 5\AA), showing that for considerably small length of the Fermi arcs, the differential conductance reduces, but still in the regime where it can be measured experimentally.

\section{Summary}
To summarize, we study the possibility of non-local transport signature of Weyl semimetal surface-states using a recursive Green's function method. We find that in presence of an applied magnetic field the surface currents flow in a Weyl semimetal in opposite directions on opposite surfaces due to their connection through the bulk states. Our study agrees with the intuition one derives from semi-classical analysis and provides the ground for further analysis and experimental observation.

\vspace{0.5cm}


\end{document}